# A New Integrated FQFD Approach for Improving Quality and Reliability of Solar Drying Systems

[a], [*] Navid Akar, [b] Hossein Lotfizadeh

[a] Department of Industrial and Systems Engineering, Northern Illinois University, DeKalb, Illinois, USA
[b] Department of Mechanical Engineering, University of Alberta, Edmonton, Alberta, Canada
[a], [*] Email: Z1725983@students.niu.edu

**Abstract**
Saffron is the most expensive spice and is significantly valuable in non-oil export. Drying process of saffron is considered as a critical control point with major effects on quality and safety parameters. A suitable drying method covering standards and market requirements while it is costlty benefitial and saves energy is desirable. Solar drying could be introduced as an appropriate procedure in rural and collecting sites of saffron since major micorobial and chemical factors of saffron can be preserved and achieved by using a renewable energy source. So, a precise system taking advantage of management, engineering and food technology sciences could be developed. Since there was no published record of integrated methods of Analytical Hierarchy Process (AHP) and Fuzzy Quality Function Deployment (FQFD) applied to solar energy drying systems, in this paper, Fuzzy Quality Function Deployment as a quality management tool by emphasizing technical and customer requirements has been implemented in order to improve quality parameters, optimizing technological expenses and market expansion strategy. Subsequently, Analytical Hierarchy Process based on survey from customers and logical pair-wise comparison are employed to decrease costs and increase the efficiency and the effectiveness of economic indicators. Using the integrated approach of AHP and FQFD in solar drying systems in saffron industry will result in cost benefit, quality improvement, the customer satisfaction enhancement, and the increase in saffron exports.

**Introduction**
Saffron spice is dehydrated by stigmas of Crocus sativus L. and is the most expensive spice. Inappropriate post-harvest processes such as poor drying process will result in reduced quality parameters. During the long periods of traditional drying enzymatic activity will be high and microbial pollution occurs [1,2]. Dehydration affects the content of compounds responsible for color, taste and aroma of saffron and also its microbial profile [1,3]. Not suitable drying may lead to scabbing, hardening of the surface, wrinkling, decline in rehydration ability, browning, surface burning and decreases in flavor and odor [4].

Agricultural yields are usually more than the immediate consumption needs, resulting in wastage of food surpluses during the short harvest periods and scarcity during post-harvest period [5]. Therefore, a reduction in the post-harvest losses of food products should have a considerable effect on the economy of these countries [6]. More than 80% of food is being produced by small farmers in developing countries [7]. These farmers dry food products by natural sun drying, an advantage being that solar energy is available free of cost, but there are several disadvantages which are responsible for degradation and poor quality of the end product. Experiments carried out in various countries have clearly shown that solar dryers can be effectively used for drying agricultural produce. It is a question of adopting it and designing the right type of solar dryer [8, 9]. Drying of agricultural produce permits; (1) early harvest; (2) planning of the harvest season; (3) long-term storage without deterioration; (4) taking advantage of a higher price a few months after harvest; (5) maintenance of the availability of seeds; and (6) selling a better quality product.

Renewable energy in terms of solar [9], wind [10], and geothermal energy [11] have been used in this industry especially in renewable based drying systems over the few decades. It is interesting to note that a large variety of drying systems that are available in the world are based on solar systems and in general, solar drying systems could be designed in direct (DSD) or indirect (ISD) types. In the former, product chamber through a clear cover receives sunlight and temperature directly. Increasing of temperature, moisture exhaust and exit by air flow will take place. In the latter system, sunlight collecting sheet, blower pump, drying chamber, side heating system and energy saving system are implemented. Parameters in designing such systems could be defined as relative moisture, the temperature of the drier, air flow rate, environment moisture and temperature, length and width of air tunnel, space and kind of trays and initial and final moisture of the product. Sun light is stored in solar collectors and then warm air flow is pumped to product chamber [12,13,14]. Drying a product over "critical moisture content" will defect significantly, however insufficient procedure will result in microbial and chemical spoilage [15]. There are a lot of researches made on drying foods as an efficient system of utilizing solar energy. Sharma et al. [16] reported solar drying of tomatoes, chilies and mushrooms. Sethi and Arora [17] improved a conventional greenhouse solar dryer of 6 m2 ×4 m2 floor area (east–west orientation) for faster drying using inclined north wall reflection (INWR) under natural as well as forced convection mode. Similar temperature behavior for drying different crops in the greenhouse drying was also presented by Jain and Tiwari [18]. A fiber reinforced plastic (FRP) hybrid solar drying house was effectively used for brown rice drying by Rachmat et al. [19]. Feasibility studies of a solar chimney to dry agricultural products were also performed by Ferreira et al. [20]. Sreekumar [21] investigated the performance of a roof-integrated solar air heating system for drying fruit and vegetables in detail by three methods namely annualized cost, present worth of annual savings, and present worth of cumulative savings. A solar tunnel dryer was used for drying grapes by Muhlbauer [22]. It was reported that the dryer produced high quality dried grapes up to the desired moisture content level. Sreekumar et al. [23] developed a new type of efficient forced convection indirect solar dryer, particularly meant for drying vegetables and fruit. In this dryer, the product was loaded beneath the absorber plate, which prevented the problem of discoloration due to irradiation by direct sunlight. Sopian et al. [24] developed and tested a double pass photovoltaic thermal solar collector suitable for solar drying applications. Comparisons were made between the experimental and theoretical results and close agreement between these two values were obtained. Singh and Kumar [25] presented convective heat transfer correlations in terms of dimensionless numbers for various common designs of solar dryer operated in different modes of air circulation. El-Beltagy et al. [26] studied Solar drying characteristics of different shapes of strawberry. The best fit of the thin layer drying of strawberry was obtained by exponential equation or Newton model which fitted very well the experimental data at various shapes of strawberry indicated by high value of coefficient correlation (r) in the range of 0.97–0.98. Akhondi et al. [27] investigated the drying of saffron stigma in a laboratory infrared dryer. The influence of temperature on the drying rate of samples at various temperatures (60, 70 . . . 110 °C) was studied. The drying time decreased with an increase in drying air temperature. According to results, the Midilli and Kucuk model adequately described the drying behavior of saffron stigmas at a controlled temperature range 60–110 °C in an infrared dryer.

Many types of solar dryers have been developed during the last two decades [28-34]; however the applications of these dryers are still limited, mainly due to their unreliable performance and high investment cost relative to a production capacity [35]. Although Mamlook has used applied fuzzy sets programing to perform evaluation of solar systems in Jordan [36], there was no significant attempt to use integrated approach of AHP and FQFD in solar technology, specifically in solar drying systems for saffron. Thus, the aim of this study is to employ this integrated method to an indirect solar dryer used for saffron to enhance consumer satisfaction. AHP technique which is expanded by Saaty [37] considers a complex system into hierarchical system of elements. It has been used in enormous researches [38-44] for making decision about multi-criteria factors which considers problems in a real situation. In reality, by applying AHP, pair-wise comparisons are made of the elements of each hierarchy by means of a normal value.

Applied QFD technique in this paper which is based on House of Quality (HOQ), includues CRs and TRs in the area of solar drying systems for saffron in order to find priority of TRs. Moreover, CRs represent customer requirements and appear as ''whats'' in the HOQ, while TRs are listed as ''hows''. Due to the need of prioritizing CRs by AHP to obtain CRs' priority weight in the process of HOQ, AHP-Fuzzy-QFD as the useful developed method has been utilized to achieve the better status of design and customer satisfaction of solar drying systems for saffron. It is worth mentioning that Fuzzy logic [45] has been also employed to deal with linguistic judgments expressing the relative importance of CRs as well as the relationships and correlations required in the HOQ.

 **Review of the Literature**
*1.1. Solar Drying System*
In this paper, an indirect solar dryer was used for the purpose of this study as shown in Fig. 1. Climatic conditions such as temperature, humidity and solar intensity and nature of the product were some major factors in designing. Multi shelf cabinet dryer was coupled with flat plate solar air heater for air heating. An air collector with air inlet was considered to collect and store the thermal energy of sunlight. Hot air is supplied through related pipe at the bottom of drying chamber and air moved upward through wire mesh of the shelves by a fan; therefore, the system operates as a forced convection system since well designed forced-convection distributed solar dryers are more effective and more controllable than natural-circulation types. The fan used for the system was a photovoltaic DC- fan. This led into having a warm air flow with a controllable temperature by a heat exchanger. The air flow reached to product chamber containing food drying trays and contact with the product. Finally, the air was exhausted from above the chamber. The dried sample was instantly transferred to laboratory for microbial and chemical tests. The solar dryer was also settled just near to laboratory towards sunlight and wind in summer season with acceptable sunlight during the day.

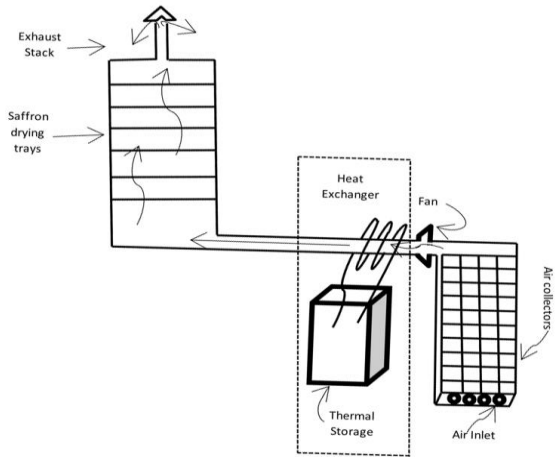

**Fig. 1.** Schematic of the indirect solar dryer used for saffron

*1.2. AHP Method*

The Analytic hierarchy process (AHP) developed by Thomas Saaty in the early 1970s is a multi-criteria decision making technique that can facilitate the decision operation by analyzing an intricate problem into a multi-level hierarchical structure of objective, criteria and alternatives [38, 40]. AHP performs pairwise evaluations to infer relative importance of the variable in each level of the hierarchy and analyzes the alternatives in the lowest level of the hierarchy in order to make the best decision among alternatives. This technique is an effective decision making method especially when subjectivity exists and it is very suitable to solve problems where the decision criteria can be organized [39, 42]. It is also used to distinguish relative priorities on absolute scales in multi-level hierarchic structures [46] (Table 1). AHP is employed in many researches, specifically in energy alternatives [47-49]. Furthermore, the proposed AHP-Fuzzy-QFD is a developed technique for determining the best criterion with respect to goal and has been used in some findings [50, 51]. Thus, in this paper, AHP as a convenient technique is employed to prioritize CRs by their priority weights in order to find important TRs in the process of fuzzy HOQ. Consequently, all of the CRs that existed in solar drying system for saffron have been considered for a hierarchical framework as a criterion to be comprised of a pairwise evaluation.

**Table 1**
Judgement scores in AHP [52]

| Option | Numerical value(s) |
| --- | --- |
| Equal | 1 |
| Marginally strong | 3 |
| Strong | 5 |
| Very strong | 7 |
| Extremely strong | 9 |
| Intermediate values to reflect inputs | 2,4,6,8 |
| Reflecting dominance of second alternatives compared with the first | Reciprocals |

*1.3. FQFD Method*

Quality function deployment (QFD) is a method to translate the customer requirements into design quality, to deploy the functions forming quality, and to employ methods for achieving the design quality aimed at satisfying the customer into subsystems and component parts, and ultimately to specific elements of the manufacturing process [53, 54]. This method was applied in 1972 in Japan for the first time as an appropriate approach to improve products quality in Japanese firms, such as Mitsubishi, Toyota, and their suppliers [55]. QFD is broadly used as an expanded quality method to satisfy customer requirements in products design of many industries recently [56-58]. In QFD most of the input variables are supposed to be precise and are considered as numerical data. However, QFD as a concept and mechanism for translating the voice of the customer into products attributes through various of product planning, engineering, processing, and production are required linguistic data to be inherently vague and ambiguity [59, 60]. Furthermore, relationships between CRs and TRs are typically imprecise and vague which is difficult to indentify them; as a result, linguistic data can be used for helping fuzzy set theory. In reality, fuzzy set theory suggests a mathematically precise way of modeling vague preferences, such as weighting performance scores on various criteria [61, 62]. It permits consideration of the different meanings that may be given to the same lingustics expression [63]. That is why this fuzzy approach has been used in several findings especifically solar systems industry.

Special attention of various subjective assessments in HOQ process is paid in 2005 by Chan and Wue for capturing the vagueness in peoples' linguistic assessments [64]. Many studies have been released by using QFD approach based on fuzzy logic to address the issue of how to deploy HOQ to improve the logistic process in various industry efficiently and effectively [65-66]. Although there is the need to link CRs and TRs with HOQ process, there is no published paper in which the authors have utilized fuzzy-QFD to enrich the design quality of producing solar drying systems in saffron industry. In the present approach, CRs and TRs in HOQ of QFD methodology are linked to recognize the most crucial technical requirements that are vital to concentrate on customer satisfaction, and optimizing the design of solar drying systems in saffron industry.

**2. AHP-Fuzzy-QFD Procedure for Solar Drying Systems of Saffron**
*2.1. Method of collecting data*

In order to collect the data, analyze and evaluate the scientific information, customer surveys and interviewed experts in the area of energy, mechanical and food engineering have been considered in this article. In reality, two types of questionaires have been utilized in this case, one for computing priority weights of customer requirements in AHP which was perfomed by customers and another for determining the relationships between CRs and TRs and the correlation between TRs in HOQ which was done by related experts. We have sent out these two distinct types of questionaires for the purpose of doing a survey of both customers and related experts efficiently as many as possible. Theorotically, the more effective experts' opinion and customers are, the more coherent and reliable the evaluation of crucial TRs in solar drying systems of saffron will be; as a result, we repeated this process for several times to reduce bias in the data and bias of one person or one voice. To facilitate the process of pair-wise comparision obtained from the survey, we designed out questionaires with ligustic variables for both customers and experts to do the pair-wise comparison in the most adequate way. Accordingly, experts and costumers responses were summarized and the further comments were elicited. Consequently, two discrete deeds were applied. 1. Entering results of the customers' survey in Expert Choice Software in order to find priority weights of customer requirements in AHP method. 2. Transforming the linquistic variables of a survey of experts to the form of fuzzy triangular numbers and analazying them in HOQ. The transforming rules from the lingustic variables to the triangular fuzzy numbers are shown in Tables 3 and 4.

*2.2. AHP-Fuzzy-QFD Framework*

The main framework of applying integrated AHP and FQFD approach to solar drying systems in saffron industry is shown in Fig. 2. Also, the fuzzy HOQ and its specific structure is illustrated in Fig. 3.

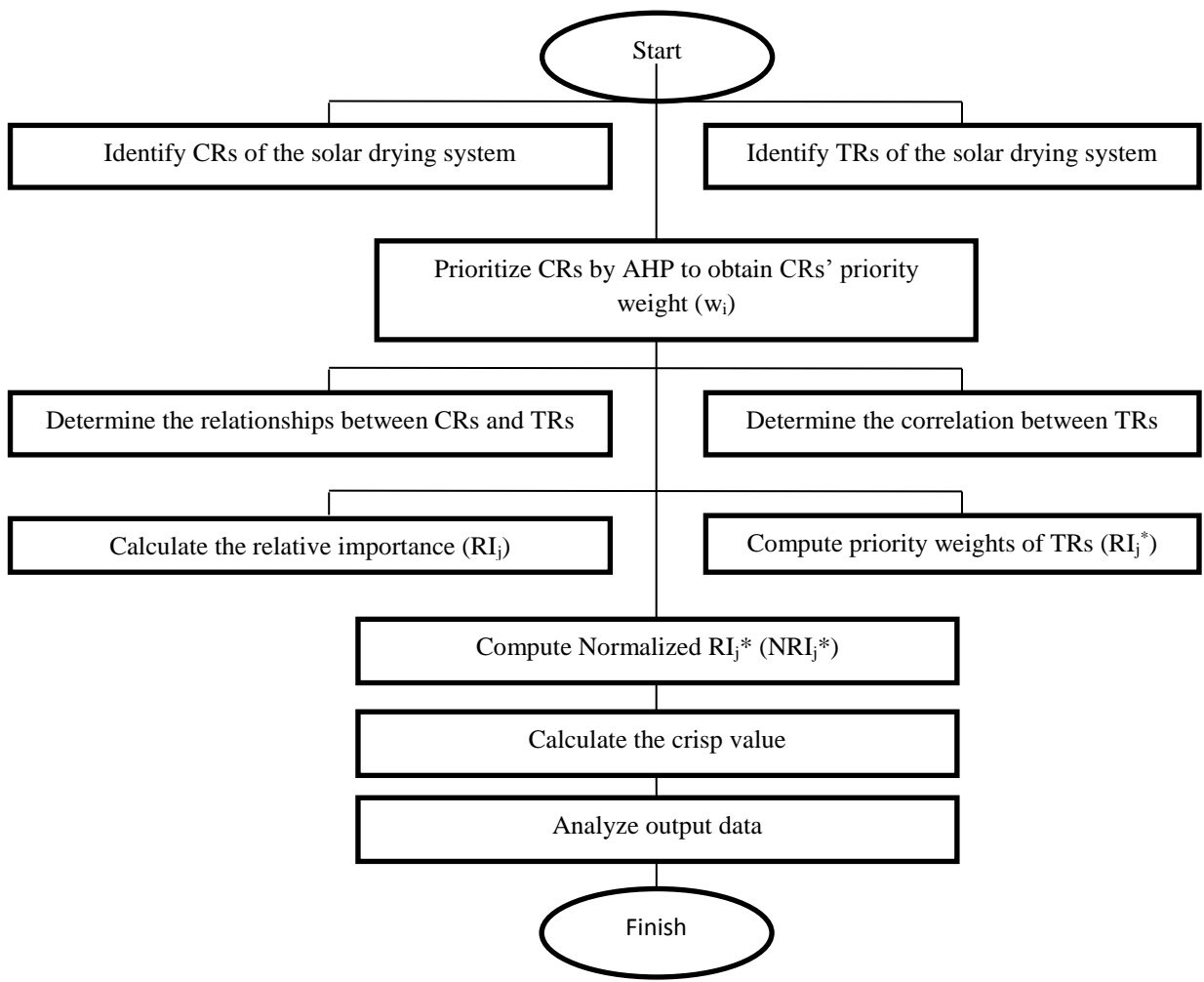

**Fig. 2.** Schematic representation of the algorithm

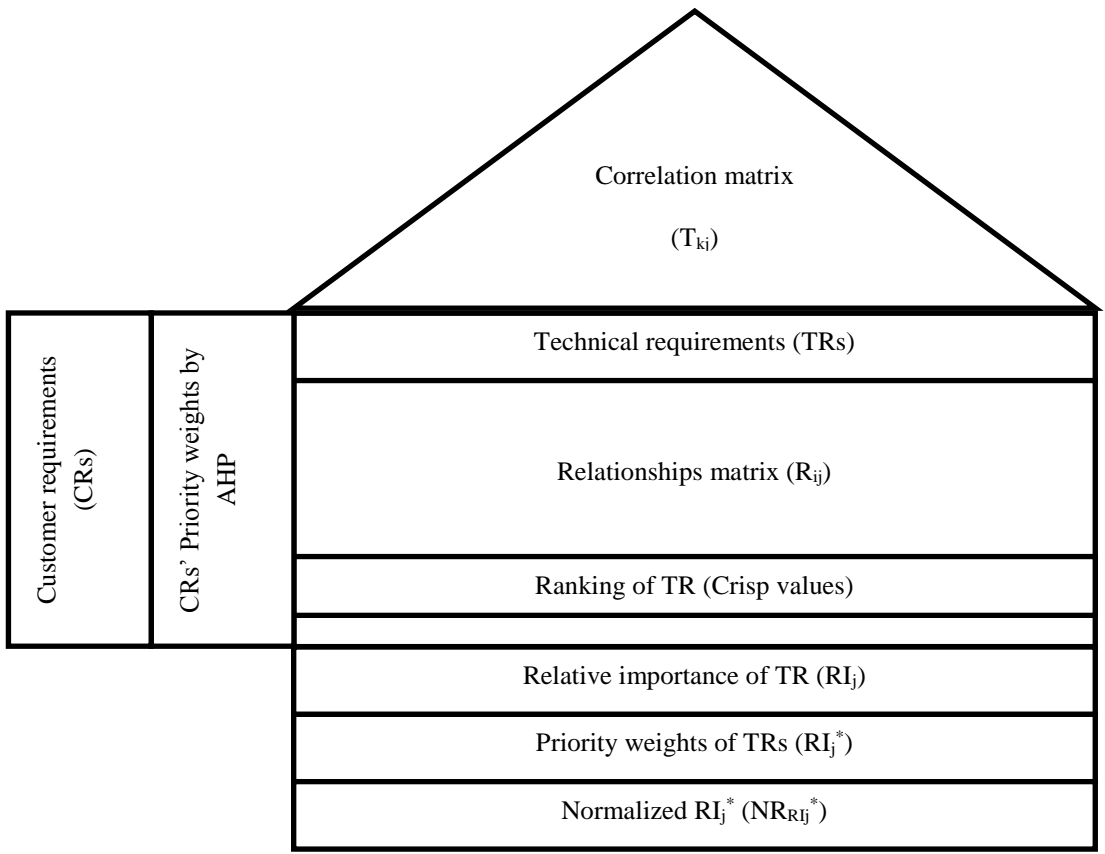

**Fig. 3.** Integration of AHP and FQFD approach

*2.2.1. Identify CRs and TRs of the solar drying system in saffron industry*

Basically, a great success of solar drying systems for saffron in a competitive market depends of course on identifying the customer requirements and providing their needs. By studying CRs and TRs gathered from the survey in the area of solar drying systems in saffron industry, fundamental requirements are considered to apply AHP-Fuzzy-QFD methodology to real cases. In actuality, CRs are defined as factors which present the underlying principles of initial steps for proceeding satisfaction of solar drying system of saffron consumers.

Accordingly, TRs are crucial features of various technologies to successfully increase customers' satisfaction. CRs present a list of customer requirements and TRs are identified in order to achieve the CRs represented in Table 2.

**Table 2**
Customer requirements and technical requirements defined for developing solar drying systems in saffron industry.

| Code of CRs | Customer Requirements | Code of TRs | Technical Requirements |
|---|---|---|---|
| $CR_1$ | Durability | $TR_1$ | Controlled conditions of temperature, humidity and time |
| $CR_2$ | Energy saving | $TR_2$ | Materials used for the air collector |
| $CR_3$ | Low required space | $TR_3$ | Collector area |
| $CR_4$ | High efficiency | $TR_4$ | Absorber coating |
| $CR_5$ | Cost-effectiveness | $TR_5$ | The length and width of the air duct |
| $CR_6$ | User-friendly | $TR_6$ | Thermal energy gained from solar radiation |
| $CR_7$ | Usable for different amount | $TR_7$ | Photovoltaic cell as the power source of the fan(s) |
| $CR_8$ | Useable for similar products | $TR_8$ | Material of the trays and their distances |
| $CR_9$ | Using alternative energy | $TR_9$ | Complete gasket and no influence on air pollution and dust |
| $CR_{10}$ | Hygiene | $TR_{10}$ | Materials used for the cabinet dryer for the heat transfer enhancement |
| $CR_{11}$ | Organoleptic properties | $TR_{11}$ | Heat insulation to enhance heat transfer |
| $CR_{12}$ | Easy Installation | $TR_{12}$ | Air temperature and humidity |
| $CR_{13}$ | Easy Portability | $TR_{13}$ | Air flow rate & velocity |
| $CR_{14}$ | Drying time | $TR_{14}$ | Calibration of thermometers and timers |

*2.2.2. Prioritize CRs by AHP to obtain CRs' priority weight (wi)*

In this step, after determining CRs, their priority weights were calculated by using Expert Choice 9.5 Software [67]. Expert Choice as a decision-making software which is based on multi-criteria decision making, implements the AHP method efficiently [68]. Moreover, it has been used in fields such as manufacturing, [43] environmental management [69, 70] and agriculture [44]. For this aim, the pairwise evaluations were prepared to analyze fourteen CRs by nine scales with respect to experimental criterions such as cost, services, and improving product quality. Finally, the obtained priority weights are considered in HOQ to extract important TRs.

*2.2.3. Determining the relationships between CRs and TRs, and the correlation between TRs*

Due to the need of a translation of imprecise and vague linguistic terms of relative importance of CRs, relationships, and correlation matrices to numerical values, fuzzy logic is utilized. In this step, the degree of relationship between TRs was then expressed by Triangle Fuzzy Numbers (TFNs) in the fuzzy HOQ. Both of these correspondences are presented in Tables 3 and 4.

As it is observed on Fig. 4, TFNs are considered which are denoted as a triplet (a, b, c) and non-fuzzy number [71, 72]:

$$\mu_N(x): \begin{cases} (x-a)/(b-a), & x \in [a,b] \\ (c-x)/(c-b), & x \in [b,c] \\ 0 & otherwise \end{cases} \quad (1)$$

**Table 3**
Degree of relationships, and corresponding fuzzy numbers (Adapted from [65])

| Degree of relationships | Fuzzy number |
|---|---|
| Strong (S) | (0.7; 1; 1) |
| Medium (M) | (0.3; 0.5; 0.7) |
| Weak (W) | (0 ; 0; 0.3) |

**Table 4**
Degree of correlations and corresponding fuzzy numbers (Adapted from [65])

| Degree of correlation | Fuzzy number |
|---|---|
| Positive (★) | (0.5; 0.7; 1) |
| Negative (▲) | (0; 0.3; 0.5) |

**Fig. 4.** Triangle fuzzy number (TFN)

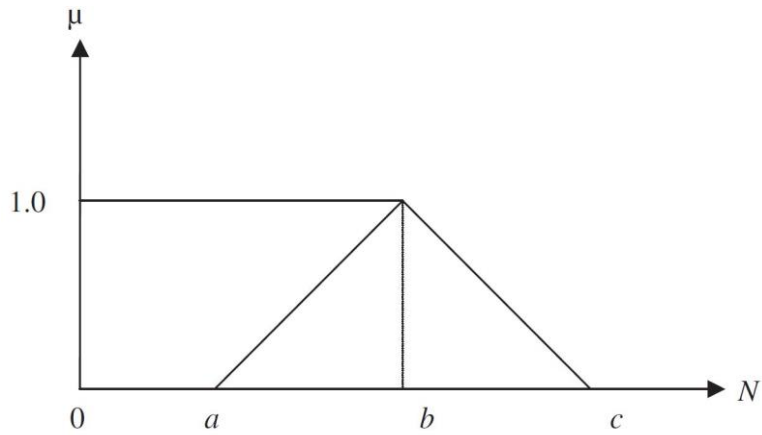

If M= ($a_1$, $b_1$, $c_1$) and N= ($a_2$, $b_2$, $c_2$) symbolize two Triangle Fuzzy Numbers, then the required fuzzy calculations are executed below [73]:

Fuzzy addition: $\quad M \otimes N = (a_1 + a_2, b_1 + b_2, c_1 + c_2)$ (2)

Fuzzy multiplication: $\quad M \otimes N = (a_1 \times a_2, b_1 \times b_2, c_1 \times c_2)$ (3)

$\quad M \otimes 1/N = (a_1/c_2, b_1/b_2, c_1/a_2)$ (4)

Fuzzy and natural number multiplication: $r \otimes M = (r.a, r.b, r.c)$ (5)

### 2.2.4. Calculating the relative importance (RIj) and priority weights of TRs (RIj*)

The aim of calculating these two parameters was to determine which TR has the most influence on developing solar drying systems in saffron industry $RI_j$ was calculated by fuzzy multiplication of $W_i$ to $R_{ij}$

$RI_j = \sum_{i=1}^{n} W_i \otimes R_{ij} \qquad j = 1, \dots, m$ (6)

$RI_j^* = RI_j \oplus \sum_{k=j} T_{kj} \otimes RI_K \qquad j = 1, \dots, m$ (7)

### 2.2.5. Calculating the normalized RIj* (NRIj*) and Crisp Value

Normalization is performed by dividing each $RI_j^*$ by the highest one according to the fuzzy set algebra [72]. Then, in order to rank the TRs, the normalized scores of $RI_j^*$ are de-fuzzified. Suppose M (a, b, c) is a Triangle Fuzzy Number; then, the crisp values are calculated using the following equation.

$\frac{(a+4b+c)}{6}$ (8)

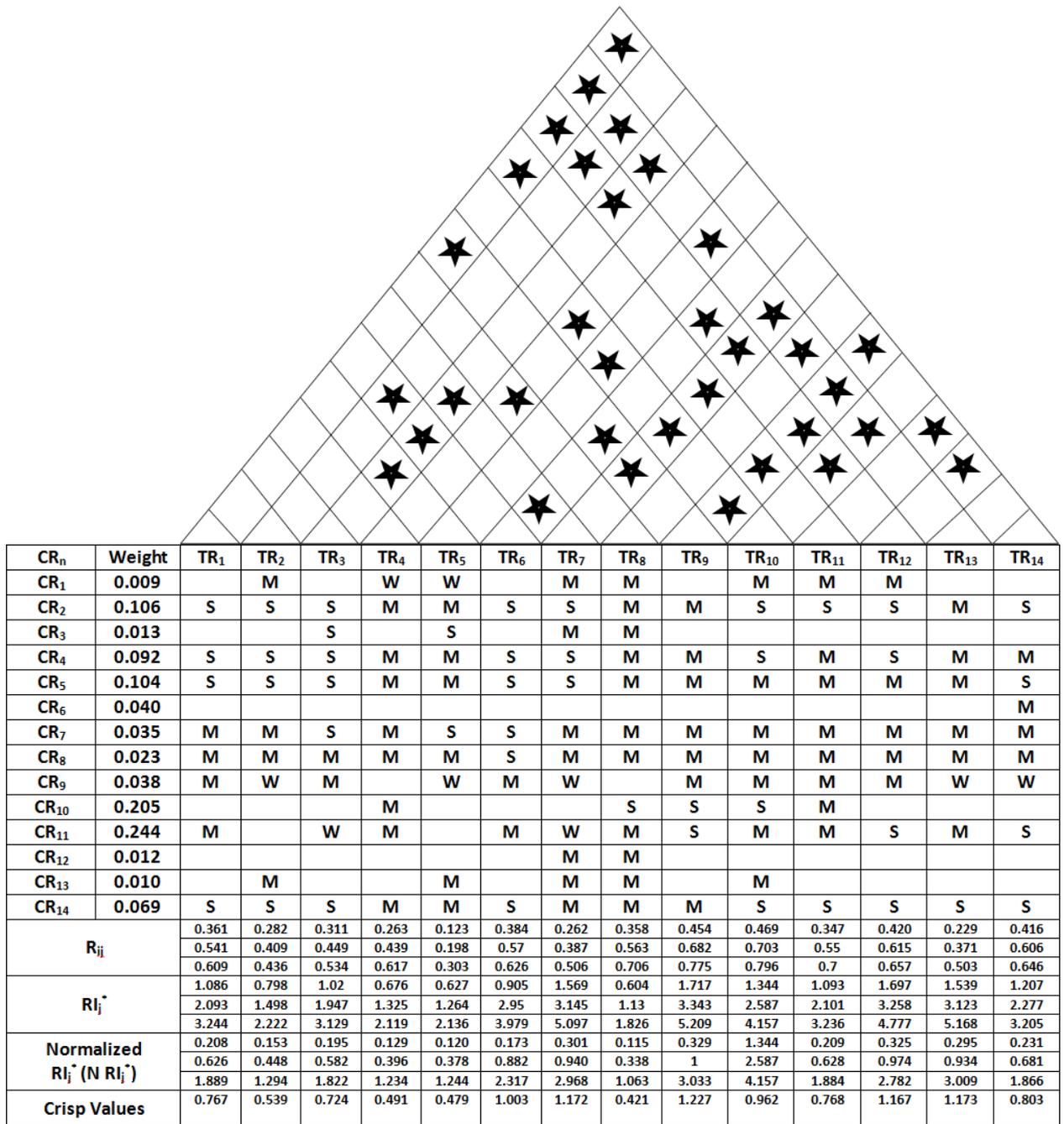

| CR$_n$ | Weight | TR$_1$ | TR$_2$ | TR$_3$ | TR$_4$ | TR$_5$ | TR$_6$ | TR$_7$ | TR$_8$ | TR$_9$ | TR$_{10}$ | TR$_{11}$ | TR$_{12}$ | TR$_{13}$ | TR$_{14}$ |
|---|---|---|---|---|---|---|---|---|---|---|---|---|---|---|---|
| CR$_1$ | 0.009 | | M | | W | W | | M | M | | M | M | M | | |
| CR$_2$ | 0.106 | S | S | S | M | M | S | S | M | M | S | S | S | M | S |
| CR$_3$ | 0.013 | | | S | | S | | M | M | | | | | | |
| CR$_4$ | 0.092 | S | S | S | M | M | S | S | M | M | S | M | S | M | M |
| CR$_5$ | 0.104 | S | S | S | M | M | S | S | M | M | M | M | M | M | S |
| CR$_6$ | 0.040 | | | | | | | | | | | | | | M |
| CR$_7$ | 0.035 | M | M | S | M | S | S | M | M | M | M | M | M | M | M |
| CR$_8$ | 0.023 | M | M | M | M | M | S | M | M | M | M | M | M | M | M |
| CR$_9$ | 0.038 | M | W | M | | W | M | W | | M | M | M | M | W | W |
| CR$_{10}$ | 0.205 | | | | M | | | | S | S | S | M | | | |
| CR$_{11}$ | 0.244 | M | | W | M | | M | W | M | S | M | M | S | M | S |
| CR$_{12}$ | 0.012 | | | | | | | M | M | | | | | | |
| CR$_{13}$ | 0.010 | | M | | | M | | M | M | | M | | | | |
| CR$_{14}$ | 0.069 | S | S | S | M | M | S | M | M | M | S | S | S | S | S |
| R$_{ij}$ | | 0.361 | 0.282 | 0.311 | 0.263 | 0.123 | 0.384 | 0.262 | 0.358 | 0.454 | 0.469 | 0.347 | 0.420 | 0.229 | 0.416 |
| | | 0.541 | 0.409 | 0.449 | 0.439 | 0.198 | 0.57 | 0.387 | 0.563 | 0.682 | 0.703 | 0.55 | 0.615 | 0.371 | 0.606 |
| | | 0.609 | 0.436 | 0.534 | 0.617 | 0.303 | 0.626 | 0.506 | 0.706 | 0.775 | 0.796 | 0.7 | 0.657 | 0.503 | 0.646 |
| RI$_j^*$ | | 1.086 | 0.798 | 1.02 | 0.676 | 0.627 | 0.905 | 1.569 | 0.604 | 1.717 | 1.344 | 1.093 | 1.697 | 1.539 | 1.207 |
| | | 2.093 | 1.498 | 1.947 | 1.325 | 1.264 | 2.95 | 3.145 | 1.13 | 3.343 | 2.587 | 2.101 | 3.258 | 3.123 | 2.277 |
| | | 3.244 | 2.222 | 3.129 | 2.119 | 2.136 | 3.979 | 5.097 | 1.826 | 5.209 | 4.157 | 3.236 | 4.777 | 5.168 | 3.205 |
| Normalized RI$_j^*$ (N RI$_j^*$) | | 0.208 | 0.153 | 0.195 | 0.129 | 0.120 | 0.173 | 0.301 | 0.115 | 0.329 | 1.344 | 0.209 | 0.325 | 0.295 | 0.231 |
| | | 0.626 | 0.448 | 0.582 | 0.396 | 0.378 | 0.882 | 0.940 | 0.338 | 1 | 2.587 | 0.628 | 0.974 | 0.934 | 0.681 |
| | | 1.889 | 1.294 | 1.822 | 1.234 | 1.244 | 2.317 | 2.968 | 1.063 | 3.033 | 4.157 | 1.884 | 2.782 | 3.009 | 1.866 |
| Crisp Values | | 0.767 | 0.539 | 0.724 | 0.491 | 0.479 | 1.003 | 1.172 | 0.421 | 1.227 | 0.962 | 0.768 | 1.167 | 1.173 | 0.803 |

**Fig. 5.** Fuzzy-HOQ for solar drying systems in saffron industry

### 3. Analyzing the output data

According to the output of Fig. 5, TRs with high crisp values indicate that they can be usefully employed to enhance relevant CRs. Therefore, the priority of TRs must be considered for developing solar drying systems for saffron. Moreover, the analyzed output of AHP-Fuzzy-QFD process gives histograms recognizing priority of TRs in Fig. 6.

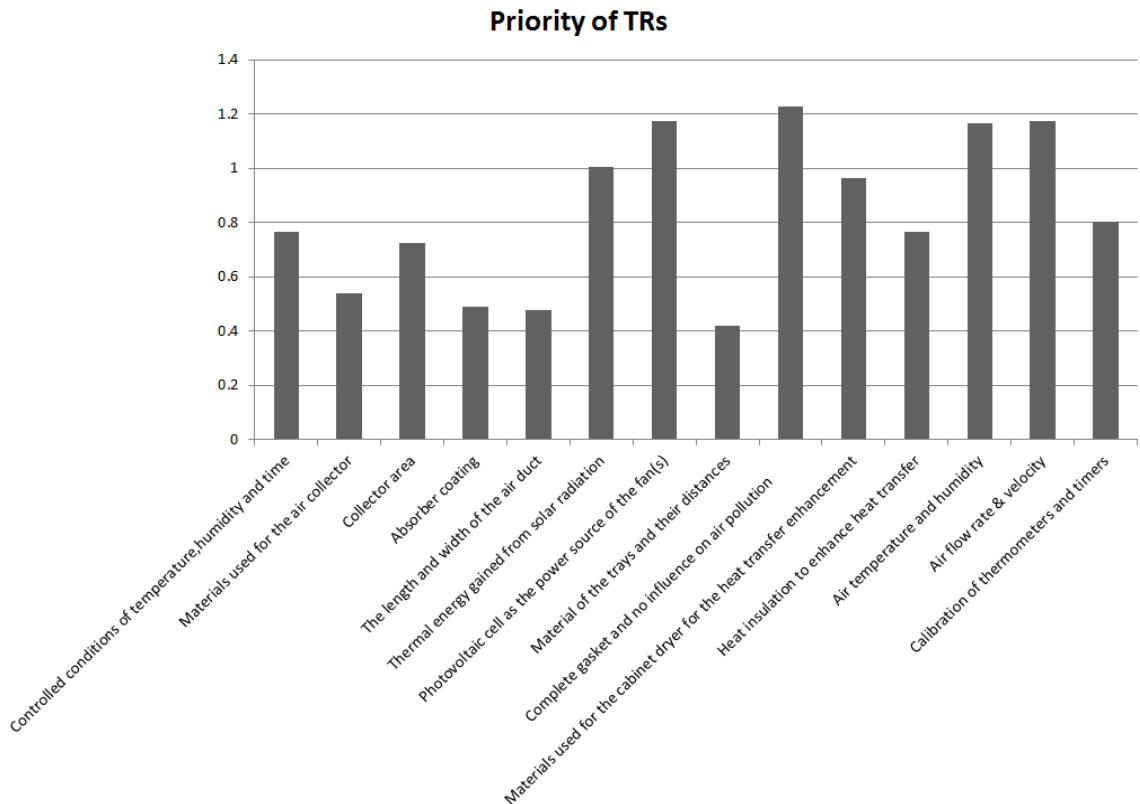

**Fig. 6.** Priority of TRs depending on crisp values

The drying behavior of agricultural crops during solar drying depends on different weather conditions including air temperature and humidity, air flow rate, and air velocity. As it can be seen in Fig. 6, after the rate of complete gasket and no influence on air pollution (1.22), the rate of air flow rate and velocity, air temperature and humidity, and photovoltaic cell as the power source of the fan (s) are virtually the same (1.18). The higher temperature, movement of the air, and lower humidity, increases the rate of drying process. If energy savings are considered, the best way to change the air flow rate of a fan is to vary the speed of rotation. To improve TRs shown on Fig. 6, there are different ways in accordance with the advancement of technology. Some of the solutions are noted in this paper. It is obvious that materials are really indispensable for having a complete gasket which has the highest rate of TRs depicted in Fig. 6 and heat transfer enhancement. For instance, using nanocoating can be a useful method to enhance heat transfer in absorber coating which can have a positive effect on the drying process. Also, suitable materials used for the cabinet dryer, air duct, heat insulation, air collector and trays are really important to increase the solar dryer efficiency. Solar dryers can be constructed from locally available and low cost materials by considering different elements related to the rate of drying. Therefore, it should be noted that materials used in solar dryers play a crucial role in drying behavior. Besides, since photovoltaic cell is utilized as the power source of the fan, a suitable photovoltaic cell should be considered to convert the energy in light into electrical energy through the process of photovoltaic. There are many types of solar cell technologies which are in development, but some of them are most commonly used such as crystalline silicon, thin films concentrators, and thermo-photovoltaic solar cell technologies. The effect of the photovoltaic cell on the drying process as a useful factor should not be neglected.

## 4. Conclusion

It is clear that drying process has major effects on the microbial profile and also the content of chemical compounds effective in color, taste and aroma of saffron. Moreover, solar energy as an unlimited renewable energy is a natural resource which has the minorest effects on microbial profile of saffron while drying process. Since, no specific published record of using integrated approach of AHP and fuzzy-QFD applied to solar drying systems for saffron have been found already, in this research, by linking CRs and TRs in fuzzy HOQ of mentioned systems we achieved the priority of TRs which should be considered in the process of design to enhance quality level of solar drying systems in saffron.

According to findings of the current study, signficance of each technical requirement by prioritizing them, has been presented in Fig. 5. The most important customer requirements such as organoleptic properties and hyigen with the weights of 0.244 and 0.205 are achieved respectively. Similarly, extracted TRs depending on crisp value from fuzzy HOQ have been depicted in Fig. 6 to present the highest priority of different TRs which allow designers to focus on them in order to increase customers satisfaction. Accordingly, the most crucial TR is the complete gasket and no influence on air pollution and dust with crisp value 1.227. Similarly, the lowest priority of TRs refer to material of the trays and their distances with crisp value 0.421. Consequently, technical suggestions for TRs developing quality and customer satisfaction of solar drying systems in saffron industry have been presented.

In this paper, using the inegrated AHP and FQFD approach indicates some advantages such as; Economical method of drying, development and employment in rural and remote areas, the development of environmentally compatible methods, intensifying the level of competitive price of saffron in global market, and developing methods of using renewable energy systems in other areas of agricultural and horticultural industries.

# Appendix

1. Nomenclature

| **Nomenclature** | | | |
|---|---|---|---|
| A | Smallest possible value | $R_{ij}$ | Relationship between the i-th CR and the j-th TR |
| B | Most promising value | $RI_j$ | Relative importance of a TR |
| C | Largest possible value | $RI_j^*$ | Priority weight of a TR |
| M and N | Triangular fuzzy numbers (TFNs) | $RI_k$ | Relative importance of the k-th TR |
| $W_i$ | Priority weight of a CR | $T_{kj}$ | Degree of correlation between the k-th and j-th TRs |